# High−Spin States and Level Structure in Stable Nucleus $^{84}$Sr

Shuifa Shen[1 2 3, *]   Xuzhong Kang[1]   Guangbing Han[5]   Shuxian Wen[6]   Yupeng Yan[2 4]   Shiwei Yan[7]

Xiaoguang Wu[6]   Lihua Zhu[8]   Guangsheng Li[6]   Chuangye He[6]

1 Engineering Research Center of Nuclear Technology Application (East China Institute of Technology), Ministry of Education, Nanchang 330013, Jiangxi, People's Republic of China

2 School of Physics, Suranaree University of Technology, Nakhon Ratchasima 30000, Thailand

3 State Key Laboratory of Nuclear Physics and Technology (Peking University), Beijing 100871, People's Republic of China

4 Thailand Center of Excellence in Physics (ThEP), Commission on Higher Education, 328 Si Ayutthaya Road, Ratchathewi, Bangkok 10400, Thailand

5 School of Physics, Shandong University, Jinan 250100, People's Republic of China

6 China Institute of Atomic Energy, P. O. Box 275(10), Beijing 102413, People's Republic of China

7 College of Nuclear Science and Technology, Beijing Normal University, Beijing 100875, People's Republic of China

8 School of Physics and Nuclear Energy Engineering, Beihang University, Beijing 100191, People's Republic of China

Abstract: High−spin states of $^{84}$Sr were populated through the reaction $^{70}$Zn($^{18}$O, 4n)$^{84}$Sr at 75 MeV beam energy. Measurement of excitation function, γ−γ coincidences, directional correlation of oriented states (DCO) ratios and γ−transition intensities were performed using eight anticompton HPGe detectors and one planar HPGe detector. Based on the measured results, a new level scheme of $^{84}$Sr was established in which 12 new states and nearly 30 new γ−transitions were identified in the present work. The positive−parity states of the new level scheme were compared with results from calculations in the framework of the projected shell model (PSM). One negative−parity band was extended to spin $I^{\pi} = 19^-$ and it can be found that in the high−spin states the γ−transition energies show the nature of signature staggering. The negative−parity band levels are in good agreement with the deformed configuration−mixing shell model (DCM) calculations.

*E-mail address: shfshen@ecit.cn

1. Introduction

Nuclei with Z≈40 and N≈45 lie in a transitional region between the deformed nuclei and spherical nuclei. There exist collective bands in these nuclei, also their structure have single−particle features, such as some isotopes of nuclei Sr, Zr and Mo etc. The neutron−deficient strontium isotopes such as $^{80}$Sr have been experimentally found to be among the most−deformed nuclei in this region. On the other hand, $^{88}$Sr with a closed neutron shell at N=50 and a fairly good proton−subshell closure at Z=38, is found to be doubly magic and, hence, spherical. $^{84}$Sr, which has four neutron holes away from $^{88}$Sr, should display both collective and quasiparticle excitations [1]; this makes it an interesting object of investigation. Dewald et al. [1], Kucharska et al. [2], Lister et al. [3], and Chattopadhyay et al. [4] have investigated the structure of $^{84}$Sr successively. About ten years ago, low−spin states of $^{84}$Sr have been studied by Döring et al. [5] whose decay data suggested that the low−lying structures of $^{80,84}$Sr show many vibration−like features in a potential with modest deformation including candidates for two− and three−phonon multiplets. This vibration−like nature seems to evolve to a more rotational behavior with increasing angular momentum $I$ and decreasing neutron number $N$. So in order to study the structure feature of $^{84}$Sr in detail, the present experiment was designed to extend the level scheme of $^{84}$Sr to higher spin states. We then calculated the positive−parity states of the new level scheme in the framework of the projected shell model (PSM). Finally, the properties of the negative−parity bands of $^{84}$Sr were also discussed qualitatively.

2. Experimental procedure

The high−spin states of $^{84}$Sr were populated through the reaction $^{70}$Zn($^{18}$O, 4n)$^{84}$Sr at a projectile energy of 75 MeV. The $^{18}$O beam was provided by HI−13 Tandem Accelerator at China Institute of Atomic Energy (CIAE). In order to increase the reaction yields in the experiment, the target consisted of a stack of two thin self-supporting $^{70}$Zn foils with a thickness of about 400 μg/cm$^2$ each. The $^{70}$Zn was isotopically enriched to 99.3%. At first the excitation function was measured using beam energies of 67, 70, 75, 80, and 87MeV, respectively. Then $\gamma-\gamma-t$ coincidence measurements were performed at optimal beam energy of 75 MeV (with beam intensity of I=8-9nA) using an array comprised of eight high-purity germanium (HPGe) detectors each with a BGO anti-Compton shield and one planar HPGe detector; each BGO(AC)HPGe detector has an efficiency of approximately 25% and the planar HPGe detector 20%. Here, *t* refers to the relative time difference between any two coincident γ rays detected within ±50 ns. In the present work, in the analysis of the data, a time-random background matrix was subtracted. These detectors were placed at angles of 35°, 36°, 39°, 43°, 81°, 90°, 102°, 143°, and 159° with respect to the beam direction, respectively, so that the directional correlation of oriented states (DCO) ratios could be deduced. The energy resolutions of the Ge detectors are between 1.8 and 2.2 keV at 1.333 MeV γ-ray energy. A total of $1.8\times10^8$ double- or higher-fold coincidence events were collected on the magnetic tapes in an event-by-event mode in the present experiment. The trigger condition for data acquisition was set such that events were recorded when at least two γ rays in the Ge detectors were in prompt coincidence. The overall count rates collected by this array during the experiment are 500-600 events/s. The γ-ray energies and relative efficiencies were calibrated with $^{133}$Ba and $^{152}$Eu sources. In order to determine the multipolarity of the γ-ray transitions, three detectors near 90° (at 81°, 90° and 102°) with respect to the beam axis were sorted against the other five detectors near 38°(at 35°, 36°, 39°, 43° and 143°) to produce a two-dimensional angular correlation matrix from which it was possible to extract the average directional correlation of oriented state (DCO) intensity ratios. The γ-γ coincidence data were analyzed with the Radware software package [6]. The interpretation of DCO ratios is most straightforward when gating is carried out on stretched E2 transitions. In this case DCO ratios of about 1.0 and 0.5 are expected for stretched ΔI=2 and ΔI=1 transitions, respectively. Consequently, the inverse value of 2.0 is expected for a quadrupole transition gated on a dipole transition, and it is certainly 1.0 for a dipole transition gated on a dipole transition. If dipole-quadrupole mixing is included, then the DCO ratio for a ΔI=1 transition may vary between 0.2 and 1.8 depending on the amount of mixing and the nuclear alignment. Ambiguities may also occur since an unstretched pure ΔI=0 transition is expected to have a ratio slightly larger than 1, i.e., similar to a stretched E2 transition.

## 3. Level scheme of $^{84}$Sr

All events were sorted off–line into a 4096×4096 disc stored matrix on which the energy windows were set later. According to already existing level scheme, at first we gated on the strongest 793 keV γ–ray, all of the relevant γ–transitions given in Ref. [4] were observed. Then gated on each of these relevant γ–transitions, the cascade relationship revealed in Ref. [4] was confirmed. In the coincidence spectrum gated by 793 keV γ–transition, we also observed the relatively strong 706, 759, and 988 keV γ–transitions, etc., and these γ–transitions are clearly visible in the coincidence spectrum gated by 1001 keV γ–transition, as shown in Fig. 1(a). In order to confirm these γ–transitions belonging to the negative–parity band of $^{84}$Sr, we gated on 988 keV γ–transition, clearly observed 510, 793, 974, and 1001 keV γ–transitions, etc., which have already been confirmed to belong to $^{84}$Sr, but can not observe the other already confirmed γ–transitions of the positive–parity band. In addition, we also observed the relatively strong 759 and 1041 keV γ–transitions, etc., in the coincidence spectrum gated by 988 keV γ–ray. In the coincidence spectrum gated by 510 keV, we observed the relatively strong 759, 988, and 1041 keV γ–transitions, etc., besides 208 and 417 keV γ–transitions. It could be seen that 759, 988, and 1041 keV γ–transitions belong to $^{84}$Sr and their cascade relationship with 510 keV γ–transition of the negative–parity band was confirmed. In the coincidence spectra gated by 759 and 719 keV, we observed relatively strong 780 keV γ–transition. From the fact that the sum of 719 and 780 equals the sum of 510 and 988, the cascade order of 988 and 759 keV γ–transitions can be obtained. At the same time, we observed 624 and 417 keV γ–transitions in the coincidence spectrum gated by 759 keV γ–transition, so their cascade order was also assigned. In the coincidence spectrum gated by 719 keV γ–transition, we observed the strong 162 keV γ–transition, and then gated on 162 keV γ–transition, proved it has a cascade relationship with 706, 719, 1088 keV γ–transitions. So far, the cascade relationship of strong γ–transitions observed in the coincidence spectrum gated by 793 keV is almost all assigned. In particular, the 706 keV γ–transition must be mentioned here. We could observe the strong 706 keV γ–transition in the coincidence spectra gated by 524 keV γ–ray of the positive–parity band and 1001 keV γ–ray, respectively. The 706 keV γ–ray can also be observed in the coincidence spectrum gated by 624 keV γ–ray, and its intensity is much stronger than that of 162 and 1088 keV γ–rays, etc., in this spectrum. These indicate that the 706 keV γ–ray exists in two different cascades. The placements of 706 keV γ–transitions were confirmed through analyzing the coincidence spectrum gated by itself, shown in Fig. 1(b). In addition, we observed the strong 964 keV γ–transition and a string of cascade transitions that are established above it in the coincidence spectrum gated by 808 keV γ–transition, shown in Fig. 1(c).

In the coincidence spectrum gated by 524 keV γ–transition of the positive–parity band, we not only observed 1202, 1205, and 1418 keV γ–transitions, but also observed 1414 and 1442 keV γ–transitions, moreover the intensity of 1414 keV γ–transition is stronger than that of 1442 keV γ–transition, and this intensity relationship also appeared in the coincidence spectra gated by 1040 and 1116 keV γ–transitions, respectively; this proved that there exists a cascade relationship between the 1414 keV and 1442 keV γ–transition. Fig. 1(d) shows the high–energy cascade part of coincidence spectrum gated by 1414 keV. In the coincidence spectrum gated by 1086 keV, we not only observed 325, 1085, 1094, and 1274 keV γ–transitions which Lister et al. [3] have already presented, but also observed 1450, 1752 and 2017 keV γ–rays, shown in Fig. 1(e). In the coincidence spectra gated by 1268 and 1418 keV γ–transitions, we also observed 1450 and 1752 keV γ–transitions, thus confirming their cascade order. In summary, 12 new levels and nearly 30 new γ–transitions were identified in the present work. Due to the weak intensities of the 1752 and 2017 keV γ–transitions, we used dashed lines to denote them in the level scheme proposed in this experiment.

As has been mentioned above, in order to assign the multipolarity of these γ–transitions and then the spins and parities of the relevant states, the detectors were positioned at nearby 38° (at 35°, 36°, 39°, 43°, and 143° with respect to the beam direction) and at nearby 90° (at 81°, 90°, and 102° with respect to the beam direction), so that the DCO ratios could be deduced. We in general gated on the strong E2 γ–transitions when the DCO ratios are extracted. In order to get better statistics, we also gated on M1 (E1) γ–transitions (e.g. 625 and 706 keV) in the analysis. The DCO ratios deduced from the present work were basically consistent with the multipolarity carried out by Chattopadhyay et al. [4]. Table 1 only lists DCO ratios of the new γ–transitions measured in the present experiment and the spins and parities of the corresponding initial and final states. Table 2 lists the relative intensities of a part of γ–transitions obtained by using the computer code SPAN98 (The relative intensities of the γ–rays were normalized to the intensity of the 793 keV transition which was set to $I_\gamma$=100). In order to reduce the interference of nearby γ–rays, i.e., to improve accuracy, we obtained the intensity of 974 keV γ–transition relative to 793 keV γ–transition in the total coincidence projected spectrum at first, and then obtained the intensities of other γ–transitions relative to 974 keV γ–transition in the coincidence spectrum gated by 793 keV γ–transition, at last the relative intensities of all these γ–transitions were normalized to the intensity of the 793 keV transition which was set to $I_\gamma$=100. The γ–transitions whose relative intensities were less than 2% were not listed in Table 2. Based on the above result, a partial level scheme of $^{84}$Sr, including the results of the present and previous works, is given in Fig. 2.

## 4. Calculations in the framework of the projected shell model

We first review briefly the main framework of the projected shell model (PSM) [7] and subsequently compare the calculated results using this model with the experimental results. The PSM is the spherical shell model built on a deformed basis. The PSM calculation usually begins with the deformed Nilsson single–particle states at a deformation $\varepsilon_2$. Pairing correlations are incorporated into the Nilsson states by BCS calculations. The consequence of the Nilsson–BCS calculations provide us with a set of quasi–particle (qp) states that define the qp vacuum $|\Phi(\varepsilon_2)\rangle$. One then constructs the shell model basis by building multi–qp states. The broken symmetry in these states is recovered by angular momentum projection [7] (and particle number projection, if necessary) to form a shell model basis in the laboratory frame. Finally a two–body shell model Hamiltonian is diagonalized in the projected space.

Following the spirit of the Tamm–Dancoff method [8], we build the shell model space by including 0−, 2− and 4−qp states for this even−even nucleus:

$$|\Phi_k\rangle = \left\{ |0\rangle, \alpha_{n_i}^+ \alpha_{n_j}^+ |0\rangle, \alpha_{p_i}^+ \alpha_{p_j}^+ |0\rangle, \alpha_{n_i}^+ \alpha_{n_j}^+ \alpha_{p_i}^+ \alpha_{p_j}^+ |0\rangle \right\}, \quad (1)$$

where $\alpha^+$ is the creation operator for a qp and index n(p) denotes neutron (proton) Nilsson quantum numbers which run over the low−lying orbitals. The basis in Eq. (1) is not arbitrarily selected but is adopted from all the neutron and proton Nilsson orbitals that lie close to the Fermi surface. Thus, the projected multi−qp states are the building blocks of the shell model basis:

$$|\psi_M^I\rangle = \sum_k f_k^I \hat{P}_{MK}^I |\Phi_k\rangle. \quad (2)$$

Here, $k$ labels the basis states and $f_k^I$ are determined by configuration mixing. We then diagonalize the Hamiltonian in the projected multi−qp states given by Eq. (2). In the calculations, we employ a quadrupole plus pairing Hamiltonian, with inclusion of quadrupole−pairing term [7].

$$\hat{H} = \hat{H}_0 - \frac{1}{2}\chi \sum_\mu \hat{Q}_\mu^+ \hat{Q}_\mu - G_M \hat{P}^+ \hat{P} - G_Q \sum_\mu \hat{P}_\mu^+ \hat{P}_\mu, \quad (3)$$

where $\hat{H}_0$ is the spherical single−particle Hamiltonian which contains a proper spin−orbit force. The other terms in Eq. (3) are quadrupole – quadrupole, monopole − and quadrupole − pairing interactions, respectively. The quadrupole − quadrupole interaction strength χ is adjusted such that the input deformation parameter ε$_2$ should be equal to the one resulting from the HFB calculation [7]. The monopole pairing strength constant G$_M$ is adjusted to give the known energy gap

$$G_M = \left[ 20.12 \mp 13.13 \frac{N-Z}{A} \right] \cdot A^{-1}, \quad (4)$$

where "−" is for neutrons and "+" for protons. Finally the quadrupole pairing strength $G_Q$ is simply assumed to be proportional to G$_M$

$$\left(\frac{G_Q}{G_M}\right)_n = \left(\frac{G_Q}{G_M}\right)_P = \gamma. \quad (5)$$

The proportionality constant $\gamma$ is fixed to be 0.2 in the present calculations. In our calculations, the pairing gaps Δ$_p$ and Δ$_n$ are calculated using the following fourth−order

expressions [9]:

$$\Delta_n = -\frac{1}{8}[M(Z,N+2) - 4M(Z,N+1) + 6M(Z,N) - 4M(Z,N-1) + M(Z,N-2)], \quad (6)$$

$$\Delta_p = -\frac{1}{8}[M(Z+2,N) - 4M(Z+1,N) + 6M(Z,N) - 4M(Z-1,N) + M(Z-2,N)]. \quad (7)$$

Where the masses of the relevant atoms, M, are taken from Ref. [10], and experimental data are adopted wherever available. The results we obtain are $\Delta_n = 1.595$ MeV and $\Delta_p = 1.83375$ MeV. The spin−orbit force parameters, κ and μ, appearing in the Nilsson potential are taken from the compilation of Zhang et al. [11] which is a modified version of Bengtsson and Ragnarsson [12] that had been fitted to the latest experimental data then. It is supposed to be applicable over a sufficiently wide range of shells. About ten years ago, based on available experimental data then, a new set of Nilsson parameters was proposed by Sun et al. [13] for proton−rich nuclei with proton (neutron) numbers 28≤N(P)≤40. Considering that the nucleus studied in the present work has a neutron number 46, we believe that this new set of parameters is not very suitable for the nucleus $^{84}$Sr, although the Nilsson parameter set proposed by Zhang et al. [11] was deduced for A≈120−140. The values κ and μ are different for different major shells (N−dependent). The calculations are performed by considering three major shells (N=2, 3, and 4) for both neutrons and protons.

In the following calculations, we construct the shell model basis at deformation parameter $β_2$=0.21 proposed by Chattopadhyay et al. [4], and assume axial symmetry although a triaxial deformation parameter γ≈8° at above $12^+$ state of $^{84}$Sr was deduced by them. It should be pointed out here that the relationship between the deformation parameters $ε_2$ and $β_2$ is the same as that in Ref. [14]. If the first term is adopted only, then $ε_2$=0.198 equals approximately $β_2$=0.21. The hexadecapole deformation parameter $ε_4$=0 is taken from the compilation of Möller et al. [10]. In the calculation of the positive−parity states of $^{84}$Sr, the configuration space is constructed by selecting the qp states close to the Fermi energy in the N=4 (N=4) major shell for neutrons (protons) and forming multi−qp states from them. Dewald et al. [1] have proposed that the two $8^+$ states of the positive−parity bands are attributed to the alignment of two−neutron and two−proton, respectively. So we assumed that the band 1 denoted in Fig. 3 predominantly has a two−neutrons ($Ω^π = +7/2^+$ and $−5/2^+$) aligned character, whereas in the band 2 the states from $8^+$ to $12^+$ predominantly have a two–protons ($\Omega^\pi = +3/2^+$ and $-1/2^+$) aligned character and from $14^+$ to $24^+$ predominantly have the 2n–2p ($\Omega^\pi_n = +7/2^+$ and $-5/2^+$, $\Omega^\pi_p = +3/2^+$ and $-1/2^+$) aligned character. The comparison of the experimentally observed positive–parity states of $^{84}$Sr with the prediction of the PSM is given in Fig. 3. As can be seen, the data is described roughly well. Due to the use of the simple BCS vacuum which only contains the properties of the ground–state rotational band, but not those of the collective vibrations [15], the calculated levels are higher than the experimental values. A microscopic description of transitional nuclei has always been challenging. In addition, the inclusion of the triaxial deformation parameter γ [4] will lower the level energies [16].

## 5. Discussion

The level scheme was compared with its neighboring even–even isotopes $^{78, 80, 82, 86}$Sr. We found that N=45 and N=46 divide the structures of the nuclei with Z=38 into two kinds of different properties, i.e., the structures of the nuclei with N<45 exhibit rotational behavior, whereas the nuclei with N>46 show strong vibrational behavior, see Refs. [17-20] for details. As can be seen from the Fig. 2, the ground–state band of $^{84}$Sr showed a certain degree of collectivity just as its neighboring isotones. Dewald et al. [1] have proposed that two $8^+$ states of the positive–parity bands attribute to the alignment of two–neutrons and two–protons, respectively, as already mentioned above. Kucharska et al. [2] and Chattopadhyay et al. [4] confirmed this analysis through the g–factor measurements. We observed γ–transitions between the close lying similar spin states of these two bands, demonstrates the existence of neutron–proton interaction, which is in agreement with the results of Chattopadhyay et al. [4]. All these are in good agreement with the calculation results carried out by Dewald et al. based upon interacting boson model (IBM) [1] and those of Chattopadhyay et al. [4]. When the spins and parities of $^{84}$Sr reached above $14^+$, it displays the properties of the rotational structure of the deformed nuclei. The signature staggering for the even–spin and odd–spin members of this band above $14^+$ can be observed. It can be seen that it no longer maintains the shape of the ground state, i.e., the shape change occurred and then the rotation happened.

It was worthy to mention that on the top of $14^+$ $^{84}$Sr still exhibited dominance of E2 γ–transitions, whereas $^{88}$Mo and $^{86}$Zr transferred to the strong M1 γ–transitions [4].

The negative–parity states were extended to an energy of 10487 keV. The first negative–parity band consisting of $5^-$, $7^-$, $9^-$, and $11^-$ levels was not extended to very high, we found only one 984 keV M1 γ–transition feeding the $11^-$ state in this band. We took 414, 432, and 624keV M1 γ–transitions which Chattopadhyay et al. [4] have already assigned through the lifetime measurements as a part of a band based on the second $5^-$ level. The $6^-$, $7^-$, and $8^-$ states which lie above the second $5^-$ have already been discovered by Dewald et al. [1] and Chattopadhyay et al. [4], and assigned their spins and parities. We extended it to spin of $19^-$ and energy of 10487 keV. It shows that the $5^-$, $6^-$, $7^-$, and $8^-$ states indeed exhibit rotational behavior. This is consistent with the prediction of Dewald et al. [1]. But the M1 γ–transitions between the $10^-$, $9^-$, and $8^-$ states were not obvious; this phenomenon is consistent with the results by means of the deformed configuration–mixing shell model (DCM) calculations in the work of Sahu [21], who suspect the prediction made by Dewald et al. [1]. However the states exhibited a very good collective vibration–like nature when the spins and parities of states are above $11^-$. In addition, we found that above $14^-$ state the γ–transition energies show the nature of signature staggering. As can be seen from the Fig. 4, the levels of the negative–parity bands proposed in the present work are in good agreement with DCM theory calculations, but the theoretical calculations haven't given higher spin states. It should be noted that the B(M1)/B(E2) ratios show an abrupt increase above the $11^-$ state, e.g. B(780)/B(988)=0.38 $(\mu_N/eb)^2$, compared with B(624)/B(1041)=28.24 $(\mu_N/eb)^2$. The observed increase in the B(M1)/B(E2) values in the negative–parity band can probably be attributed to neutron holes occupying the $g_{9/2}$ high–$\Omega$ orbitals [4].

## 6. Summary

In summary, the high–spin states of the nucleus $^{84}$Sr have been investigated by means of conventional in–beam γ–ray spectroscopy. Twelve new levels and nearly thirty new γ–transitions have been added to the old level scheme. One negative–parity band was extended to spin $I^\pi = 19^-$. The positive–parity states of the level scheme have been discussed in the framework of the projected shell model. The calculated results also demonstrate that $^{84}$Sr shows vibration – like feature. Finally, the properties of the negative – parity band were also discussed.

We would like to thank the operating staff of the HI – 13 Tandem Accelerator at China Institute of Atomic Energy for providing the $^{18}$O beam. The project is supported by Suranaree University of Technology under contract No. 15/2553, the Major State Basic Research Development Program in China under Contract No. 2007CB815003, the National Natural Science Foundation of China under Grant Nos. 10547140, 11065001, 11075214, 10927507, 10975191, 10675171 and 61067001.

## References

[1] A Dewald, U Kaup, W Cast et al., Phys. Rev. C 25(1982)226.

[2] A.I. Kucharska, J. Billowes, C.J. Lister, J. Phys. G 15(1989)1039.

[3] C.J. Lister, P. Chowdhury, D. Vretenar, Nucl. Phys. A 557(1993)361c.

[4] S. Chattopadhyay, H.C. Jain, M.L. Jhingan et al., Phys. Rev. C 50(1994)93.

[5] J. Döring, A. Aprahamian, M. Wiescher, J. Res. Natl. Inst. Stand. Technol. 105(2000)43.

[6] D. C. Radford, Nucl. Instrum. Methods Phys. Res. A 361, 297 (1995).

[7] K. Hara and Y. Sun, Int. J. Mod. Phys. E 4(1995)637.

[8] P. Ring and P. Schuck, The Nuclear Many – Body Problem (Springer – Verlag, New York, 1980)282 – 285.

[9] P. Möller and J. R. Nix, Nucl. Phys. A536(1992)20.

[10] P. Möller, J. R. Nix, W. D. Myers and W. J. Swiatecki, At. Data Nucl. Data Tables 59(1995)185.

[11] J.–Y. Zhang, N. Xu, D. B. Fossan, Y. Liang, R. Ma and E.S. Paul, Phys. Rev. C 39(1989)714.

[12] T. Bengtsson and I. Ragnarsson, Nucl. Phys. A 436(1985)14.

[13] Y. Sun, J.-Y. Zhang, M. Guidry, J. Meng, and S. Im, Phys. Rev. C 62, 021601 (2000).

[14] R. Bengtsson, J. Dudek, W. Nazarewicz, and P. Olanders, Phys. Scr. 39, 196 (1989).

[15] Y. Sun and C. L. Wu, Phys. Rev. C 68(2003)024315.

[16] Xuzhong Kang, Shuifa Shen, Jianzhong Gu, Yupeng Yan, Xiaoguang Wu, Lihua Zhu, Shiwei Yan, and Tingdun Wen, J. Phys. Soc. Jpn. 80, 044201 (2011).

[17] D. Rudolph, C. Baktash, C.J. Gross et al., Phys. Rev. C 56(1997)98.

[18] B. Singh, Nucl. Data. Sheets. 66(1992)639-641.

[19] H.W. Müller, Nucl. Data. Sheets. 50(1987)17.

[20] C.A. Fieds, F.W.N. Deboer, J. Sau et al., Nucl. Phys. A 398(1983)512.

[21] R. Sahu, Nucl. Phys. A 501(1989)311.

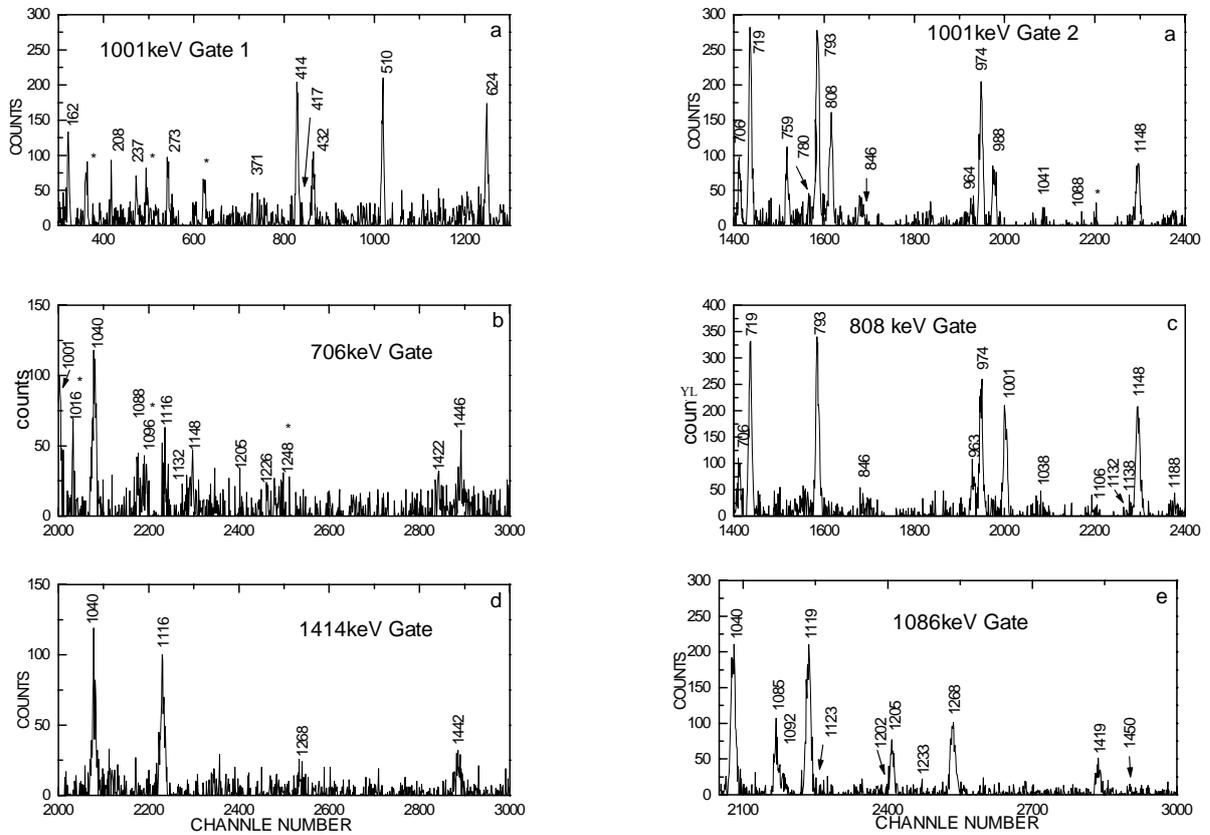

Fig. 1. Some examples of coincidence spectra gated by (a) 1001, (b) 706, (c) 808, (d) 1414, and (e) 1086 keV $\gamma$−rays, respectively. The asterisk denotes the ray that does not belong to $^{84}$Sr.

Fig. 2. The partial level scheme of $^{84}$Sr proposed in the present work. The transition energies are given in units of keV.

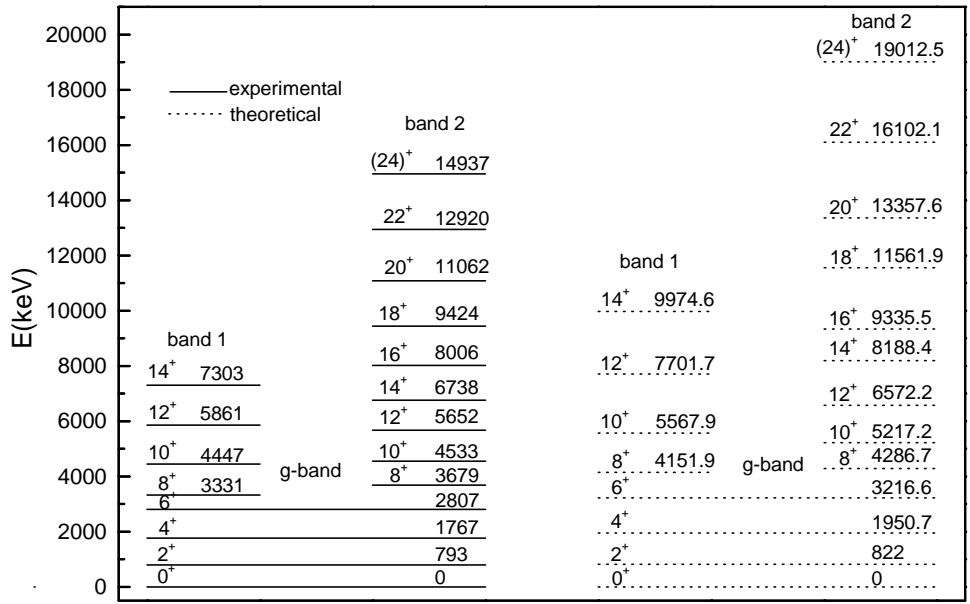

Fig. 3. Projected shell model calculations for positive–parity states in $^{84}$Sr compared with the experimental data.

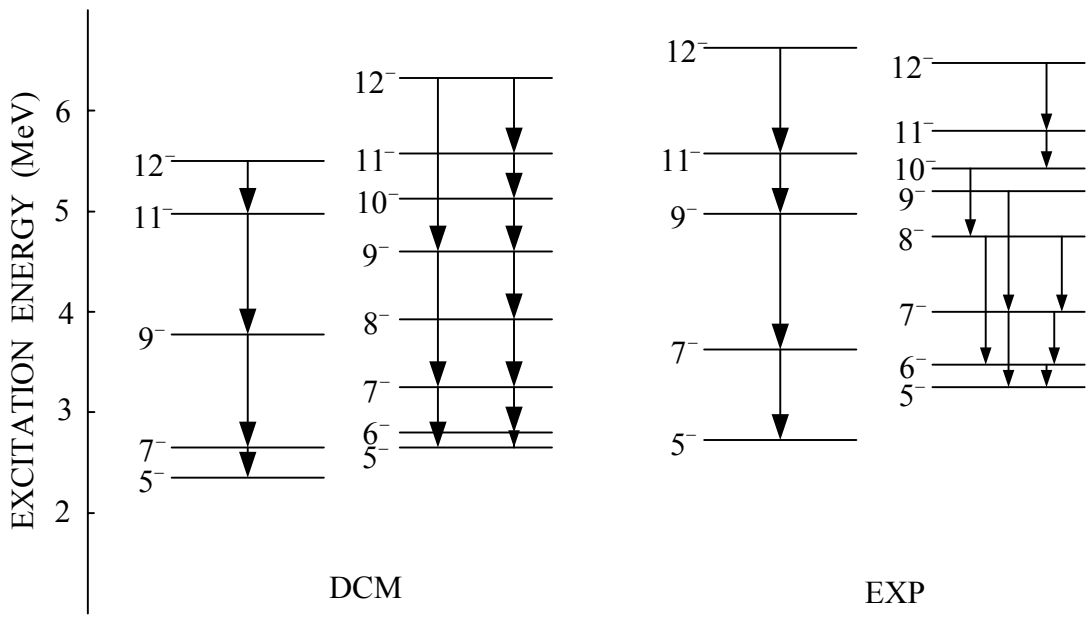

Fig. 4. DCM calculations for negative–parity states in $^{84}$Sr compared with the present experimental data.

Table 1. Some new γ–transitions of $^{84}$Sr, the corresponding initial level energies, and their DCO ratios deduced from the present experiment.

| $E_x$(keV) | $E_\gamma$(keV) | $I_i$ | $I_f$ | DCO | $E_x$(keV) | $E_\gamma$(keV) | $I_i$ | $I_f$ | DCO |
|---|---|---|---|---|---|---|---|---|---|
| 5861 | 1414 | $12^+$ | $10^+$ | | 3278 | 510 | $6^-$ | $5^-$ | $0.36^{d)}$ |
| 7303 | 1442 | $14^+$ | $12^+$ | | 5025 | 759 | $10^-$ | $8^-$ | $0.95^{c)}$ |
| 4533 | 1202 | $10^+$ | $8^+$ | $0.91^{e)}$ | 6067 | 1041 | $12^-$ | $10^-$ | $1.07^{c)}$ |
| 7823 | 1085 | $15^+$ | $14^+$ | $0.69^{e)}$ | 6913 | 846 | $14^-$ | $12^-$ | $0.43^{b)}$ |
| 8006 | 182 | $16^+$ | $15^+$ | 0.64 | 7619 | 706 | $15^-$ | $14^-$ | $0.92^{b)}$ |
| 9097 | $1274^{[3]}$ | $17^+$ | $15^+$ | $0.76^{e)}$ | 8139 | 1226 | $16^-$ | $14^-$ | $0.10^{b)}$ |
| 9097 | 1094 | $17^+$ | $16^+$ | $0.34^{e)}$ | 7619 | 1138 | $15^-$ | $13^-$ | $0.33^{b)}$ |
| 9424 | 325 | $18^+$ | $17^+$ | | 7512 | 1030 | $14^-$ | $13^-$ | $0.94^{b)}$ |
| 10547 | 1123 | $19^+$ | $18^+$ | $0.48^{e)}$ | 8752 | 1240 | $16^-$ | $14^-$ | $0.37^{b)}$ |
| 10547 | 1450 | $19^+$ | $17^+$ | $1.00^{e)}$ | 8752 | 1132 | $16^-$ | $15^-$ | $1.08^{b)}$ |
| 11062 | 514 | $20^+$ | $19^+$ | | 9065 | 1446 | $17^-$ | $15^-$ | $0.56^{a)}$ |
| 12299 | 1752 | $21^+$ | $19^+$ | $1.00^{e)}$ | 10487 | 1422 | $19^-$ | $17^-$ | $0.69^{a)}$ |
| 12299 | 1238 | $21^+$ | $20^+$ | $0.58^{e)}$ | 7512 | 1105 | $14^-$ | $12^-$ | $0.82^{f)}$ |
| 12920 | 1858 | $22^+$ | $20^+$ | $1.05^{e, g)}$ | 4737 | 1088 | $9^-$ | $7^-$ | $0.75^{c)}$ |
| 14937 | 2017 | $(24^+)$ | $22^+$ | $1.06^{e, g)}$ | 3649 | 162 | $7^-$ | $7^-$ | $0.88^{c)}$ |
| 3487 | 208 | $7^-$ | $6^-$ | $0.66^{f)}$ | 6407 | 964 | $12^-$ | $11^-$ | $0.78^{f, h)}$ |
| 4266 | 780 | $8^-$ | $7^-$ | | 7619 | 108 | $15^-$ | $14^-$ | |
| 5443 | 417 | $11^-$ | $10^-$ | | 6481 | 74 | $13^-$ | $12^-$ | |
| 4266 | 988 | $8^-$ | $6^-$ | $0.85^{d)}$ | | | | | |

a), b), c), d), e), and f) gated by 706, 625, 719, 793, 1086, and 808keV, respectively, when DCO ratios are deduced; g) obtained from the spectrum without subtracting the background; h) obtained according to the systematics.

Table 2. The relative intensities and initial state energies of a part of γ – transitions of $^{84}$Sr.

| $E_\gamma$(keV) | 86 | 272 | 348 | 372 | 414 | 432 | 510 | 524 | 609 |
|---|---|---|---|---|---|---|---|---|---|
| $I_\gamma$ | 9.8(6) | 2.0(4) | 8.3(6) | 1.5(4) | 14.3(6) | 6.5(6) | 10.9(7) | 40.8(8) | 3.0(2) |
| $E_x$(keV) | 4533 | 3040 | 3679 | 3649 | 6481 | 6913 | 3278 | 3331 | 3649 |
| $E_\gamma$(keV) | 625 | 680 | 706 | 719 | 759 | 793 | 808 | 854 | 872 |
| $I_\gamma$ | 10.6(6) | 6.2(6) | 9.5(6) | 12.5(6) | 2.5(4) | 100 | 10.1(6) | 24.3(7) | 18.6(7) |
| $E_x$(keV) | 6067 | 3487 | 7619, 5443 | 3487 | 5025 | 793 | 5443 | 4533 | 3679 |
| $E_\gamma$(keV) | 964 | 974 | 988 | 997 | 1001 | 1040 | 1085 | 1088 | 1116 |
| $I_\gamma$ | 1.3(4) | 79.4(2) | 2.9(4) | 1.3(3) | 18.3(6) | 58.8(9) | 16.5(6) | 2.8(3) | 11.0(5) |
| $E_x$(keV) | 6407 | 1767 | 4266 | 5443 | 2768 | 2807 | 7823 | 4737 | 4447 |
| $E_\gamma$(keV) | 1119 | 1148 | 1205 | 1268 | 1442 | 1445 | | | |
| $I_\gamma$ | 14.1(5) | 5.5(4) | 2.1(4) | 0.8 | 0.8(2) | 1.4(2) | | | |
| $E_x$(keV) | 5652 | 4635 | 5652 | 8006 | 7303 | 9065 | | | |